\documentclass[11pt,twoside]{article}

\usepackage{asp2006}
\usepackage{epsf}
\usepackage{psfig}
\usepackage{lscape}

\begin{document}
\title{Detection of a Distinct Pseudobulge Hidden Inside the ``Box-Shaped Bulge'' of NGC 4565}   \author{John C. Barentine and John Kormendy}   
\affil{Department of Astronomy, University of Texas at Austin, 1 University
Station C1400, Austin, TX 78712-0259, USA}    

\begin{abstract} 
$N$-body simulations show that ``box-shaped bulges'' of edge-on galaxies are not bulges at all: they are bars seen side-on.  The two components that we readily see in edge-on Sb galaxies like NGC 4565 are a disk and a bar.  But face-on SBb galaxies always show a disk, a bar, and a (pseudo)bulge. Where is the (pseudo)bulge in NGC 4565?  We use archival Hubble Space Telescope $K$-band images and Spitzer Space Telescope 3.6 $\mu$m wavelength images to penetrate the dust in NGC 4565.  We find a high surface brightness, central stellar component that is clearly distinct from the boxy bar and from the galaxy's disk.  Its minor-axis profile has a S\'{e}rsic index of 1.33$\pm$0.12, so it is a pseudobulge.  The pseudobulge has the smallest scale height ($\sim$90) pc of any component in the galaxy.  This is in contrast to a scale height of $\sim$740 pc for the boxy bar plus thin disk. The disky pseudobulge is also much less luminous than the boxy bar, so the true pseudobulge-to-total luminosity ratio of the galaxy is much less than previously thought.  We infer that the (pseudo)bulge-to-total luminosity ratios of edge-on galaxies with box-shaped bulges have generally been overestimated.  Therefore more galaxies than we have recognized contain little or no evidence of a merger-built classical bulge.  This presents a challenge to our picture of galaxy formation by hierarchical clustering, because it is difficult to grow big galaxies without also making a big classical bulge. Solving the puzzle of the ``missing pseudobulge'' in NGC 4565 further increases our confidence that we understand box-shaped bulges correctly as edge-on bars.  This in turn supports our developing picture of the formation of pseudobulges -- both edge-on bars and disky central components -- by secular evolution in isolated galaxies. 
\end{abstract}

\section{Motivation}

\begin{figure}[htb]
\center{
\leavevmode
\epsfbox{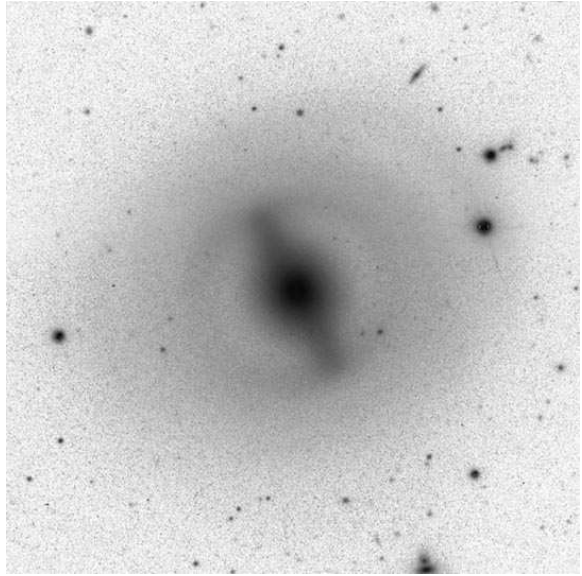}
\caption{The nearly face-on SB0 galaxy NGC 4608 in a $gri$ composite image from the Sloan Digital Sky Survey.}
}
\label{n4608}
\end{figure}

Hierarchical clustering of galaxies dominated their evolution in the early Universe but is currently yielding to internal, slow (``secular'') evolution.  The evolution of galaxies via interactions with collective phenomena like bars and spiral structure is well-established \citep{KK04}. $N$-body simulations suggest that bars heat up in the axial direction and explain the morphology of so-called ``box-shaped'' bulges of some edge-on galaxies \citep{CombesSanders81}.  These secular processes drive gas toward the galactic center, building to high densities and often triggering starbursts.  In the process, bulge-like (``pseudobulge'') structures are produced mimicking ``classical'' (i.e., merger-built) bulges.  These structures are easy to identify in nearly face-on objects such as NGC 4608, a normal, early-type barred galaxy shown in Figure~\ref{n4608}
\epsfysize=3.0in
The disk, bar and pseudobulge are clearly discernible in this image.  To what extent are these same components seen in (nearly) edge-on galaxies?

An edge-on galaxy with a box-shaped bulge we can use to check is NGC 4565.  Short wavelength imagery shows a prominent dust lane in NGC 4565 that may hide interior structures due to the shallow viewing angle.  Infrared images of this galaxy penetrate the dust and show the disk and the edge-on bar manifesting as a box-shaped bulge.  Using infrared data from the Spitzer Space Telescope and Hubble Space Telescope, we set out to (1) search for the underlying pseudobulge our picture of secular evolution tells us should exist and (2) determine the scale heights for the various components of the galaxy.

\section{Method}

\begin{figure}[htb]
\center{
\leavevmode
\epsfbox{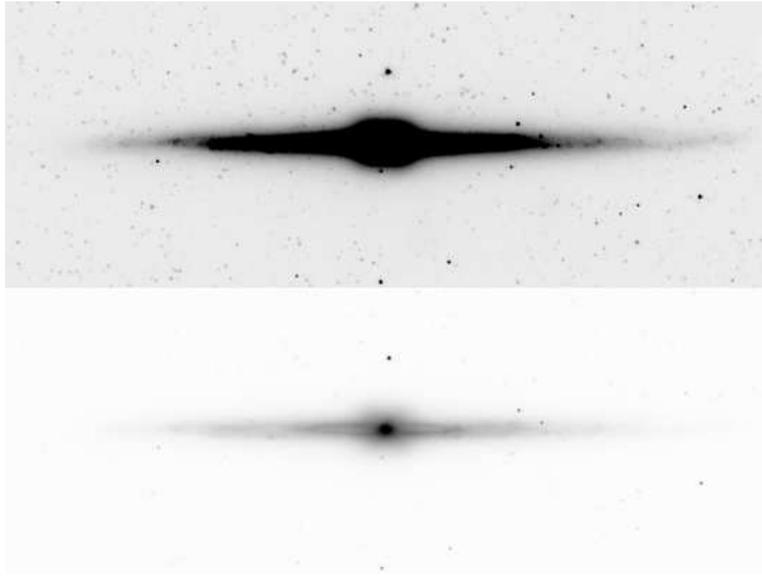}
\caption{Spitzer/IRAC 3.6 $\mu$m image of NGC 4565 presented in two different stretches, emphasizing the boxy bar (top) and  the inner ring and pseudobulge (bottom).}
}
\label{n4565}
\end{figure}

In edge-ons, extinction at optical wavelengths is large along the sightline through the highly-inclined disk.  Observations in infrared solve this problem.  We used 3.6 $\mu$m Spitzer/IRAC archive images to see through the dust and measure the minor axis light profile of NGC 4565.  The imagery is shown in Figure~\ref{n4565} with stretches emphasizing the box-shaped bulge (top) and the central pseudobulge (bottom).  We supplemented the Spitzer data at small radii with NICMOS F160W imagery from the HST archive; the higher spatial resolution of the NICMOS data allowed us to measure the S\'{e}rsic index of the pseudobulge.  Profiles were measured by taking vertical cuts along the minor axis of NGC 4565 after registering the HST and Spitzer images.  Surface photometry was done using these cuts from which we performed a three-component decomposition into a central S\'{e}rsic function for the pseudobulge, another S\'{e}rsic function for the boxy bulge, and an outer exponential.  The data and fits are shown in Figure 3.
\epsfysize=3.0in

\epsfysize=3.0in
\begin{figure}[htb]
\center{
\leavevmode
\epsfbox{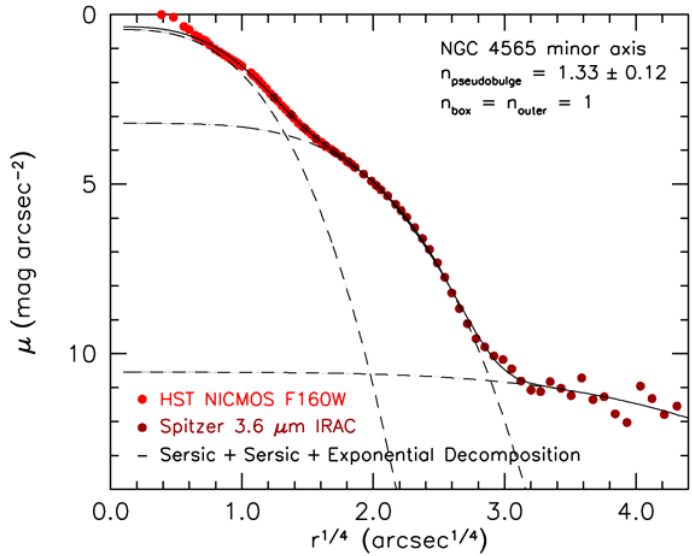}
\caption{The measured brightness profile of the center of NGC 4565 along the minor axis from HST/NIC-MOS F160W (red points) and Spitzer/IRAC 3.6 $\mu$m data (brown points).  The dashed lines show a decomposition of the profile into components in order of increasing radius: the pseudobulge (S\'{e}rsic), box-shaped bar (S\'{e}rsic), and outer disk (exponential).  The solid line is the sum of the components.}
}
\label{profile_fits}
\end{figure}

\section{Results}

We find that the innermost component of the decomposition is best fit by a profile with S\'{e}rsic index $n$ = 1.33$\pm$0.12, consistent with a pseudobulge rather than a classical bulge \citep{FisherDrory08}.  Excluding the faint thick disk and halo, the thickest parts of the galaxy are in decreasing order of thickness: the edge-on bar, the thin disk, and the pseudobulge.  The boxy bar and thin disk have a combined scale height of 10$\farcs$5 ($\sim$740 pc at a distance of 14.5 Mpc; \citet{Wu02}), while the pseudobulge scale height is only 1$\farcs$2 ($\sim$90 pc).  \citet{Wu02} fit a profile to the ``bulge'' in NGC 4565, defining this structure as what remained after subtracting fitted profiles for the disks and halo.  Their fitted scale height for the structure we interpret as a pseudobulge is 0.65 kpc.  Our value compares favorably with theirs, but they fitted an exponential rather than a S\'{e}rsic function to their optical data and claim that photometry alone cannot resolve uncertainty about whether the box structure is a bar or not.  We assert that when the ÒrealÓ bulge in NGC 4565 is revealed to be a pseudobulge, interpretation of the structures becomes more straightforward.

\citet{SimienGdV86} find the bulge-to-total light ratio ($B/T$) = 0.4 in 4565 but $B$ refers to the boxy bar -- not the pseudobulge within.  Figure~\ref{n4565} shows that the pseudobulge is clearly less luminous than the boxy structure.  If NGC 4565 were seen face-on, light from the apparent box-shaped bulge would be recognized as a bar and would lead to a lower $B/T$ value.  Previously measured $B/T$ ratios of edge-on galaxies with box-shaped bulges are probably overestimates, given this reasoning.

\section{Conclusions}

The interpretation of boxy bulges in edge-on galaxies as signature of bars is more believable if we find pseudobulges like those associated with bars in face-on galaxies.  Our discovery of the pseudobulge in NGC 4565, distinct from the box-shaped bar previously thought to be the bulge, increases confidence in our picture of secular evolution.   Furthermore, $B/T$ ratios in edge-on galaxies with boxy bulges are smaller than previously believed.  Published $B/T$ values for most edge-on galaxies must be inconsistent with those derived for their face-on counterparts.  Finally, overestimates of $B/T$ in edge-ons present a problem with respect to CDM galaxy formation.  The disk of NGC 4565 rotates at 255$\pm$10 km s$^{-1}$ interior to the outer warp \citep{Rupen91}; it has thus grown to great mass while showing no evidence for a major merger.  It is difficult to reconcile these observations in the context of a hierarchically clustering universe.

\acknowledgements 

These results are based on observations made with the Spitzer Space Telescope, which is operated by the Jet Propulsion Laboratory, California Institute of Technology under a contract with NASA.  Additional data from the NASA/ESA Hubble Space Telescope were used, obtained from the data archive at the Space Telescope Institute, operated by the association of Universities for Research in Astronomy, Inc. under the NASA contract NAS 5-26555.  This work was supported by the National Science Foundation under grant AST-0607490.  The image of NGC 4608 was furnished by David W. Hogg, Michael R. Blanton, and the Sloan Digital Sky Survey Collaboration.

\end{document}